\documentclass[12pt]{article}
\usepackage{times}
\usepackage{geometry}
\usepackage[utf8]{inputenc}
\usepackage{enumitem,amssymb}
\usepackage{ragged2e}
\usepackage{graphicx}
\usepackage{floatrow}
\newlist{thematic}{itemize}{8}
\setlist[thematic]{label=$\square$}
\usepackage{pifont}
\newcommand{\cmark}{\ding{51}}%
\newcommand{\done}{\rlap{$\square$}{\raisebox{2pt}{\large\hspace{1pt}\cmark}}%
\hspace{-2.5pt}}

\begin{document}
\raggedright
\huge
Astro2020 Science White Paper \linebreak

The role of Active Galactic Nuclei in galaxy evolution: insights from space ultraviolet spectropolarimetry \linebreak
\normalsize

\noindent \textbf{Thematic Areas:} \hspace*{60pt} $\square$ Planetary Systems \hspace*{10pt} $\square$ Star and Planet Formation \hspace*{20pt}\linebreak
$\done$ Formation and Evolution of Compact Objects \hspace*{31pt} $\square$ Cosmology and Fundamental Physics \linebreak
  $\square$  Stars and Stellar Evolution \hspace*{1pt} $\square$ Resolved Stellar Populations and their Environments \hspace*{40pt} \linebreak
  $\done$    Galaxy Evolution   \hspace*{45pt} $\square$             Multi-Messenger Astronomy and Astrophysics \hspace*{65pt} \linebreak
  
\textbf{Principal Author:}

Name: Fr\'ed\'eric Marin	
 \linebreak						
Institution: Universit\'e de Strasbourg, CNRS, Observatoire Astronomique de Strasbourg, UMR 7550, F-67000 Strasbourg, France.
 \linebreak
Email: frederic.marin@astro.unistra.fr
 \linebreak
Phone: +33 368 852 459
 \linebreak
 
\textbf{Co-authors:} S. Charlot (Sorbonne Universit\'e, UPMC-CNRS, UMR7095, Institut d'Astrophysique de Paris, F-75014, Paris, France), 
		    B. Ag\'is Gonz\'alez (Space Sciences, Technologies and Astrophysics Research (STAR) Institute, Universit\'e de Li\`ege, All\'ee du 6 Ao\^ut 19c, B5c, 4000 Li\`ege, Belgium), 
		    D. Sluse (Space Sciences, Technologies and Astrophysics Research (STAR) Institute, Universit\'e de Li\`ege, All\'ee du 6 Ao\^ut 19c, B5c, 4000 Li\`ege, Belgium), 
		    D. Hutsem\'ekers (Space Sciences, Technologies and Astrophysics Research (STAR) Institute, Universit\'e de Li\`ege, All\'ee du 6 Ao\^ut 19c, B5c, 4000 Li\`ege, Belgium), 
		    A. Labiano (Centro de Astrobiolog\'ia (CAB, CSIC-INTA), ESAC Campus, E-28692 Villanueva de la Ca\~nada, Madrid, Spain), 
		    L. Grosset (LESIA, Paris Observatory, PSL University, CNRS, Sorbonne Universit\'e, Univ. Paris Diderot, Sorbonne Paris Cit\'e, 5 place Jules Janssen, 92195 Meudon, France), 
		    C. Neiner (LESIA, Paris Observatory, PSL University, CNRS, Sorbonne Universit\'e, Univ. Paris Diderot, Sorbonne Paris Cit\'e, 5 place Jules Janssen, 92195 Meudon, France), 
		    and J.-C. Bouret (Aix Marseille Univ., CNRS, CNES, LAM, Marseille, France)
  \linebreak

\pagebreak

\textbf{Introduction} 

~\

Active galactic nuclei (AGN) are extra-galactic objects observable from the nearby Universe out to cosmological distances thanks to their unmatched brightness among non-transient 
sources (Bañados et al. 2018). The tremendous amount of photons emitted by an AGN arise from mass accretion onto a central supermassive black hole (SMBH). 
Matter falling into the SMBH potential well forms an accretion disk which converts gravitational potential energy into kinetic energy.
This kinetic energy is in turn transformed into internal energy (i.e. heat) thanks to friction processes occurring in the disk (Shakura \& Sunyaev 1973). This is an extremely
efficient way of producing radiation from the far ultraviolet (UV) to the near-infrared, with photons originating from a region not larger that a few hundred/thousand
astronomical units. AGN are also powerful sources of high-energy (X-rays and gamma-ray) radiation, thermal-infrared re-emission by dust and, 
depending on the existence of a kiloparsec-scale jet of collimated material, synchrotron radio emission (see Fig. 1). 

~\

Although the global AGN picture seems rather understood, the details of AGN physics are still poorly constrained. The key role of AGN and the growth of SMBH in galaxy evolution
is now recognized and highlighted by numerous observational studies and hydrodynamical simulations (see, e.g., Kormendy \& Richstone 1995; Kauffmann \& Haehnelt 2000; Sijacki et al. 2015).
However, the small size of the central engine (less than a parsec) makes it challenging to spatially resolve the innermost regions of an AGN and its host galaxy
with current spatial-resolution capabilities, and hence, to understand the precise mechanisms of matter and energy transfer and SMBH growth. Indeed, SMBH growth is 
sketchy but potentially large at low redshift (z $\sim$ 2, Daddi et al. 2007), while the early SMBH growth phase (at z $>$ 7) remains largely unknown (Elvis 2009). The strong radiation
fields emerging from central regions of AGN or from radio-dominant jets are almost always accompanied by kinetic energy transfer. AGN have been observed to produce high-velocity
winds at a fraction of the speed of light (e.g., Tombesi et al. 2010), which can extend over several hundred parsec before diluting into the interstellar medium.
Winds are responsible for the radiative/kinetic perturbation of the dynamics and thermal state of the local environment (Fabian 2012). The even more powerful 
jets of magnetized plasma can reach the intergalactic medium, pushing everything along their path. Thus, AGN interact tightly with their host galaxies, as reflected also
by the observed strong correlations between SMBH mass and properties of the host galaxy, such as bulge luminosity, mass and velocity
dispersion (e.g., Kormendy \& Ho 2013), and by the need for an energetic (feedback) process able to remove or heat gas in massive galaxies to prevent further star formation 
(see e.g., Kauffmann \& Haehnelt 2000). Indeed, AGN hosts are, on average, more massive, compact, centrally peaked, and pressure-supported than non-active galaxies. They 
appear to lie in a transition region between star-forming and non-star-forming galaxies (Sanchez et al. 2018), indicating that AGN are in the process of 
halting/quenching star formation, effectively transforming galaxies between different families (Silk \& Rees 1998). This brings a crucial question: how does AGN 
feedback on galaxy evolution actually work?

\begin{figure}
  \floatbox[{\capbeside\thisfloatsetup{capbesideposition={right,top},capbesidewidth=7cm}}]{figure}[\FBwidth]
  {\caption{\small Unscaled sketch of an AGN. At the centre lies a supermassive 
  black hole around which a multi-temperature accretion disc spirals 
  (shown with the colour pattern of a rainbow.). The accreted matter is
  illuminated by an X-ray corona (shown in violet) of unknown size. The 
  disk extends from $\sim$ 10$^{-5}$ pc to $\sim$ 10$^{-2}$ pc for a 
  10$^8$ solar masses black hole. The region responsible for the emission 
  of broad lines is in red and light brown. It extends out to $\sim$ 0.1 pc,
  where the circumnuclear dusty region (shown in dark brown) onsets. The
  collimated polar ionized winds (in green) are created in sub-parsec 
  scale regions and their final extension interacts with the interstellar
  medium (shown in yellow-green at a few hundred parsecs). A double-sided,
  kilo-parsec jet is added to account for jet-dominated AGN (Marin 2016).}\label{Fig1}}
  {\includegraphics[width=9cm]{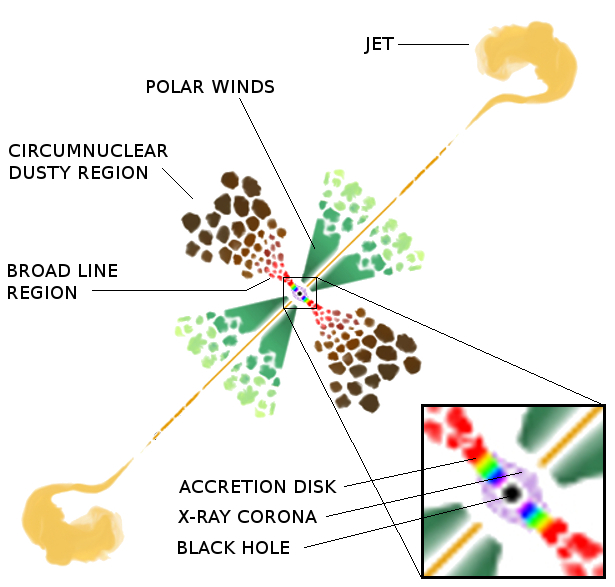}}
\end{figure}

~\

Significant progress in these fields was achieved thanks to the advent of high-resolution X-ray grating spectrometers (Chandra/LETGS, Chandra/HETGS, XMM-Newton/RGS), high-angular 
resolution imaging using either large radio array (ALMA, VLA) or adaptive optics in high-contrast (VLT, Keck), timing information (WHT/ULTRACAM, GBT/ARCONS), and interferometry 
(VLBI, VLBA). Yet, all these techniques are intrinsically limited by their lack of sensitivity to the morphology of the emitting/reprocessing regions and the intensity and configuration of 
the magnetic field. Polarimetry, on the other hand, is extremely sensitive to the geometry and magnetic-field properties of the light source. Unlike spectroscopy, which is 
limited by the physical size of the emitting region, polarimetry allows us to probe spatially-unresolved volumes and determine with precision their geometrical and magnetic topologies. 
Additionally, polarimetry provides information about matter composition, temperature, and velocity, which complements that obtainable via other techniques (most notably high-resolution 
spectroscopy). Hence, new polarimetric observations using state-of-the-art polarimeters promise us the strongest constraints on the physical parameters controlling SMBH 
growth, AGN radiative outputs and their impact on galaxy evolution.

~\

\textbf{Probing the spatially unresolved central engine with ultraviolet spectropolarimetry} 

~\

Some key signatures of accretion disks can be revealed only in polarized light, and with higher contrast at ultraviolet than at longer wavelengths. Specifically, models of disk 
atmospheres usually assume Compton scattering in an electron-filled plasma, resulting in inclination-dependent polarization signatures (up to 10\%, see e.g., Chandrasekhar 1960). 
Yet optical polarization is detected at less than a percent, and parallel to the radio jets if any (Stockman et al. 1979). Whether these low levels can be attributed to dominant 
absorption opacity (Laor \& Netzer 1989) or complete Faraday depolarization (Agol \& Blaes 1996) is unclear. This degeneracy can be broken by looking at the numerous UV lines that
are formed in the innermost AGN regions (e.g, Ly$\alpha$~$\lambda$1216, C~{\sc ii}~$\lambda$1335, C~{\sc iv}~$\lambda$1549, Mg~{\sc ii}~$\lambda$2800 ...). These lines are the key
to understanding UV polarization, and only observations with high signal-to-noise ratio and high spectral resolution can distinguish between the two effects. If absorption opacity is 
responsible for the low continuum polarization we detect, the line profiles should also show a significant drop in polarization. High-resolution spectropolarimetry is thus mandatory 
if we want to better understand the immediate vicinity of SMBH. This requires a space telescope with a large collecting area (e.g. LUVOIR) able to provide UV spectropolarimetry.

~\

At larger distances from the accretion disk, but still at (sub)parsec scale, the role and origin of the gas and dust reservoir situated along the disk plane must also
be investigated. Is matter being transferred from the host to the AGN, or is this dusty medium outflowing from the accretion disk? Interestingly, in our Galaxy,
dust extinction, which is highest in the UV, shows a local peak near 2175~\AA~(Stecher \& Donn 1965). The strength of this feature varies from galaxy to galaxy: it is 
weaker in the Large and Small Magellanic Clouds than in the Galaxy and almost never observed in AGN (Gaskell \& Benker 2007). Unveiling the mineralogy of extragalactic dust grains 
is not easy and requires high-quality extinction-curve measurements. A strong advantage for any new-generation UV polarimeter is that polarization induced by dust scattering rises
rapidly toward the blue, peaking near 3000~\AA\ in the rest frame and remaining nearly constant at shorter wavelengths (see, e.g., Hines et al. 2001). Polarimetry at short wavelengths 
can thus discriminate between dust scattering and wavelength-independent electron scattering. Information about the associated polarization position angle would also set tight 
constraints on the dust location.  Indeed, the decomposition of an AGN spectral energy distribution suggests that the mid-infrared component corresponds to equatorial
emission, approximately aligned with the plane of the inner accretion disk, while the weaker near-infrared peak might be associated with hot dust in the inner polar region. This 
picture has been challenged recently by the finding in some AGN that the bulk of infrared emission originates from the polar region above the circumnuclear dust, 
where only little dust should be present (e.g., Asmus et al. 2016). Polarimetric observations in a band little affected by starlight dilution (e.g., the UV band) should help us determine 
where the polarized signal comes from: a polarization angle parallel to the radio axis would originate primarily from an equatorial dust distribution, 
and a perpendicular polarization angle primarily from a polar dust component (Lopez-Rodriguez et al. 2018). Such finding would provide strong constraints on 
evolutionary models of AGN, the origin and geometries of dusty winds, and the energy and kinetic budgets of galactic enrichment. 

~\

Polarimetric measurements can not only probe the dust composition in extragalactic objects, but also help us determine magnetic-field topology 
and strength. Theory predicts that paramagnetic grains will be aligned with their longer axes perpendicular to the local magnetic field if exposed to magnetic or anisotropic-radiation 
fields with wavelengths less than the grain diameter (Lazarian \& Hoang 2007). Therefore, the UV band can not only selectively trace the smallest dust grains, allowing better 
characterization of AGN dust composition, but since polarization degree is predicted to be proportional to magnetic-field strength, UV polarimetry would enable first measurements
of the intensity and direction of the magnetic field on parsec scales around an AGN core (Hoang et al. 2014). Also, radiative pumping of fine-structure levels
in atoms and ions in a magnetic field is expected to give rise to polarized line emission. A number of prominent UV lines are predicted to show significant 
polarization following that mechanism, providing a means of tracing the magnetic field in hot AGN gas on sub-parsec scales (Yan \& Lazarian 2008). On larger scales, synchrotron 
polarization from prominent jets can also teach us about the evolution of the magnetic-field configuration. As an example, polarization measurements of 
Fanaroff-Riley class I AGN indicate that the component of the magnetic field projected on the plane of the sky is first parallel to the axis of the jet close to the central engine, 
and then becomes transverse at larger distances (Laing et al. 2011). The transition region, controlled by poorly constrained physical processes, can be investigated only with 
polarimetry. Finally, not only does polarization probe the magnetic-field structure of jets, serving as a hydrodynamic tracer of shocks, bends, and shear, but it also 
probes the medium through which it propagates by encoding the signature of Faraday effects along the line of sight (Homan 2005).

~\

\textbf{AGN outflows and their impact on the host galaxy} 

~\

Dissipative processes in the accretion disk transfer matter inward, angular momentum outward, and heat up the disk. Magnetic-field lines from the inner part of the accretion disk 
cross the event horizon of the black hole and are wound up by its spin, launching Poynting flux-dominated outflows (see, e.g., Blandford \& Znajek 1977). The resulting jets tend to 
be collimated for a few parsecs and dilute in giant lobes on kilo-parsec scales. Relativistic electrons traveling in ordered magnetic fields are responsible for the high polarization
we detect (of the order of 40--60\%, see e.g., Thomson et al. 1995). Interestingly, the continuum polarization degree and angle are extremely sensitive to the strength and direction 
of the magnetic field, and to the charge distribution. A high-energy polarimeter would be able to probe in detail the magnetic configuration of such jets by measuring the 
electron-beam polarization in different regions. If a jet is inclined toward the observer (blazar-like objects), a non-thermal spectral energy distribution will be observed, with a 
low-energy broadband peak in the radio-to-UV wavelength range. Comparing the observed high-energy polarization degree of blazars to leptonic, hadronic or alternative jet models will 
enable better constraints on the composition and lifetimes of particles in the plasma (Zhang 2017). Since jets are also responsible for ion and neutrino emission, they are valuable 
sources to understand how cosmic rays are produced.

~\

In addition to jets, strong outflows will be important targets for future high-resolution polarimeters. At redshifts greater than 1.5--2.0, a sub-category of AGN, called 
Broad-Absorption-Line quasars (BAL QSO), exhibit very broad absorption features in UV resonant lines (Ly$\alpha$, C~{\sc iv}, Si~{\sc iv}). The gas outflows producing these signatures
presumably contribute to the enrichment of the AGN host galaxies. BAL QSO are particularly interesting as they tend to have high polarization degrees ($>$ 1\%, e.g., Ogle et al. 1999), 
which can be used to constrain wind geometry (Young et al. 2007). These BAL QSO are believed to be the high-redshift analogues of more nearby, polar-scattered Seyfert galaxies, whose 
UV and optical emission can be explored with spatial polarimeters. High-resolution spectropolarimetry, achievable from space with a telescope with large collecting area (e.g. LUVOIR), would provide
new constraints on wind kinematics, for the first time from UV resonance lines, similarly to what has been achieved by Young et al. (2007) using the H$\alpha$ line. UV polarimetry, 
in combination with optical data, would enable the first detailed and consistent picture of the launch of outflowing winds in targets which we can 
resolve spatially (a goal unachievable with higher-redshift targets).

\begin{figure}
  \floatbox[{\capbeside\thisfloatsetup{capbesideposition={right,top},capbesidewidth=7cm}}]{figure}[\FBwidth]
  {\caption{\small FUV -- NUV colour map of the main body of NGC~7252. The pixels 
  are colour-coded in units of FUV -- NUV colour. The point spread function for the
  UV Imaging Telescope on ASTROSAT is shown in black circle. The image measures 
  $\sim$ 50" times 50" and corresponds to a physical size of $\sim$ 16 kpc on each 
  side. Age contours of 150 (red), 250 (green), 300 (blue) Myr are overlaid over the 
  colour map to isolate regions of constant age. The blue ring of star formation 
  quenching is clearly seen with bluer colour clumps. The ring hosts young ($<$ 150 Myr)
  stellar populations compared to the rest of the galaxy. Figure from 
  Georges et al. (2018).}\label{Fig2}}
  {\includegraphics[width=9cm]{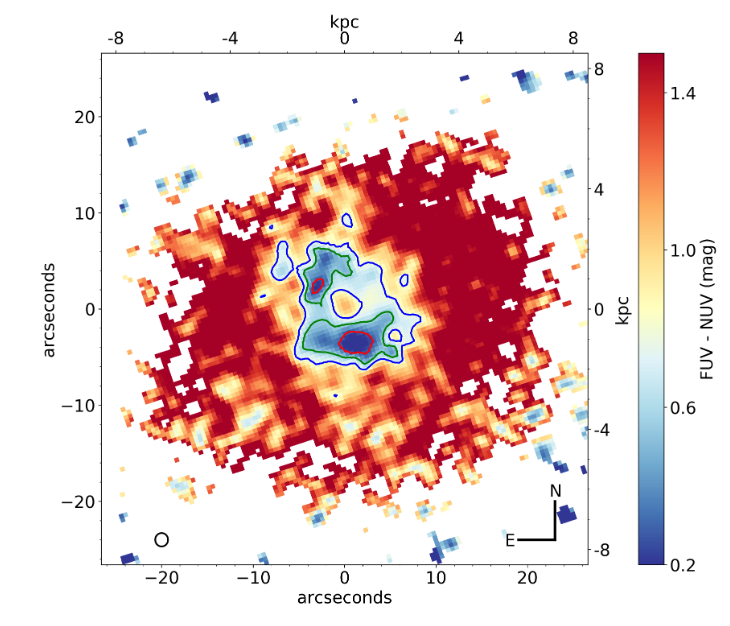}}
\end{figure}

~\

AGN also have a strong impact on the star-formation and chemical-evolution histories of their host galaxies. When an AGN episode onsets in a galaxy, 
the radiative and kinetic power transferred from the AGN to the host can easily quench, suppress or re-activate star formation, profoundly altering the whole galaxy (see Fig. 2 and
Wagner et al. 2012). The extent to which the AGN impacts on its host through this feedback is largely unknown.
Polarization is needed to quantitatively determine the AGN contribution to feedback by 
measuring the broadband synchrotron (de)polarization signatures. As summarized by Mao et al. (2014), at radio wavelengths,
minimal AGN contribution would result in polarized emission from a geometrically
thin layer located at the top of the jet lobes, while a major contribution to star formation/quenching would result in large-scale, diffuse Faraday signatures detectable 
in polarization. In addition to broadband polarimetry, high-resolution spectropolarimetry can bring valuable information about the processes acting on star formation thanks to the Zeeman effect 
associated with intense local magnetic fields. The star formation episode can be probed through spectral-line polarization, such as achieved for the starburst galaxy NGC~1808 using the H$\alpha$ line (Scarrott et al. 1993). Extended surveys of emission-line polarization compared to other AGN activity indicators across the UV/visible spectrum can help us quantify in this way the impact of AGN feedback on the evolutionary path of galaxies.

\pagebreak

\textbf{References}

~\

{\small
Agol, E., \& Blaes, O. 1996, MNRAS, 282, 965 \\
Blandford, R.~D., \& Znajek, R.~L. 1977, MNRAS, 179, 433 \\
Ba{\~n}ados, E., Venemans, B.~P., Mazzucchelli, C., et al. 2018, Nature, 553, 473 \\
Chandrasekhar, S. 1960, New York: Dover, 1960 \\
Daddi, E., Alexander, D.~M., Dickinson, M., et al. 2007, ApJ, 670, 173 \\
Elvis, M. 2009, astro2010: The Astronomy and Astrophysics Decadal Survey, 2010 \\
Fabian, A.~C. 2012, ARA\&A, 50, 455 \\
Gaskell, C.~M., \& Benker, A.~J. 2007, arXiv:0711.1013 \\
George, K., Joseph, P., Mondal, C., et al. 2018, A\&A, 613, L9 \\
Hines, D.~C., Schmidt, G.~D., Gordon, K.~D., et al. 2001, ApJ, 563, 512 \\
Hoang, T., Lazarian, A., \& Martin, P.~G. 2014, ApJ, 790, 6 \\
Homan, D.~C. 2005, Future Directions in High Resolution Astronomy, 340, 133 \\
Kauffmann, G., \& Haehnelt, M. 2000, MNRAS, 311, 576 \\
Kishimoto, M., Antonucci, R., Blaes, O., et al. 2008, Nature, 454, 492 \\
Kormendy, J., \& Richstone, D. 1995, ARA\&A, 33, 581 
Kormendy, J., \& Ho, L.~C. 2013, ARA\&A, 51, 511 \\
Laing, R.~A., Guidetti, D., Bridle, A.~H., Parma, P., \& Bondi, M. 2011, MNRAS, 417, 2789 \\
Lamy, H., \& Hutsem{\'e}kers, D. 2004, A\&A, 427, 107 \\
Laor, A., \& Netzer, H. 1989, MNRAS, 238, 897 \\
Lazarian, A., \& Hoang, T. 2007, ApJL, 669, L77 \\
Lazarian, A., \& Hoang, T. 2008, ApJL, 676, L25 \\
Lopez-Rodriguez, E., Alonso-Herrero, A., Diaz-Santos, T., et al. 2018, MNRAS, 478, 2350 \\
Mao, S.~A., Banfield, J., Gaensler, B., et al. 2014, arXiv:1401.1875 \\
Marin, F. 2016, MNRAS, 460, 3679 \\
Ogle, P.~M., Cohen, M.~H., Miller, J.~S., et al. 1999, ApJs, 125, 1 \\
S{\'a}nchez, S.~F., Avila-Reese, V., Hernandez-Toledo, H., et al. 2018, RMxAA, 54, 217 \\
Scarrott, S.~M., Draper, P.~W., Stockdale, D.~P., \& Wolstencroft, R.~D. 1993, MNRAS, 264, L7 \\
Shakura, N.~I., \& Sunyaev, R.~A. 1973, A\&A, 24, 337 \\
Sijacki, D., Vogelsberger, M., Genel, S., et al. 2015, MNRAS, 452, 575 
Silk, J., \& Rees, M.~J. 1998, A\&A, 331, L1 \\
Stecher, T.~P., \& Donn, B. 1965, ApJ, 142, 1681 \\
Stockman, H.~S., Angel, J.~R.~P., \& Miley, G.~K. 1979, ApJL, 227, L55 \\
Thomson, R.~C., Robinson, D.~R.~T., Tanvir, N.~R., Mackay, C.~D., \& Boksenberg, A. 1995, MNRAS, 275, 921 \\
Tombesi, F., Cappi, M., Reeves, J.~N., et al. 2010, A\&A, 521, A57 \\
Wagner, A.~Y., Bicknell, G.~V., \& Umemura, M. 2012, ApJ, 757, 136 \\
Young, S., Axon, D.~J., Robinson, A., Hough, J.~H., \& Smith, J.~E. 2007, Nature, 450, 74 \\
Zhang, H. 2017, Galaxies, 5, 32
}

\end{document}